\documentclass[preprint2]{aastex} 

\usepackage{amsmath}
\usepackage{natbib}

\shorttitle{Solar Convection Spectrum}
\shortauthors{Hathaway  et al.}

\begin{document}
\title{The Sun's Photospheric Convection Spectrum}
 
\author{David H. Hathaway}
\affil{NASA Ames Research Center, Moffett Field, CA 94035 USA}
\email{david.hathaway@nasa.gov}

\author{Thibaud Teil}
\affil{NASA Ames Research Center, Moffett Field, CA 94035 USA}
\email{thibaud.teil@gmail.com}

\author{Aimee A. Norton}
\affil{W.W. Hansen Experimental Physics Laboratory, Stanford University, Palo Alto, CA 94305 USA}
\email{aanorton@stanford.edu}

\author{Irina Kitiashvili}
\affil{NASA Ames Research Center, Moffett Field, CA 94035 USA}
\email{irina.n.kitiashvili@nasa.gov}

\begin{abstract}
Spectra of the cellular photospheric flows are determined from full-disk Doppler velocity observations
acquired by the Helioseismic and Magnetic Imager (HMI) instrument on the Solar Dynamics Observatory (SDO) spacecraft.
Three different analysis methods are used to separately determine spectral coefficients
representing the poloidal flows, the toroidal flows, and the radial flows.
The amplitudes of these spectral coefficients are constrained by simulated data analyzed
with the same procedures as the HMI data.
We find that the total velocity spectrum rises smoothly to a peak at
a wavenumber of about 120 (wavelength of about 35 Mm), which is typical of supergranules.
The spectrum levels off out to wavenumbers of about 400, and then rises again
to a peak at a wavenumber of about 3500 (wavelength of about 1200 km), which is typical of granules.
The velocity spectrum is dominated by the poloidal flow component (horizontal flows
with divergence but no curl) at wavenumbers above 30.
The toroidal flow component (horizontal flows with curl but no divergence) dominates
at wavenumbers less than 30.
The radial flow velocity is only about 3\% of the total flow velocity at the lowest wavenumbers,
but increases in strength to become about 50\% at wavenumbers near 4000.
The spectrum compares well with the spectrum of giant cell flows at the lowest wavenumbers and with the spectrum
of granulation from a 3D radiative-hydrodynamic simulation at the highest wavenumbers.
\end{abstract}


\section{INTRODUCTION}

The Sun is the only star for which we can directly observe the convective motions that carry energy
from its radiative interior to its photosphere.
While buoyancy and heat transport drive these motions, these flows also transport
magnetic field and angular momentum and thus play a critical role in the rotational and magnetic
history of the star.
The convective motions, and their interactions with magnetic fields, are the ultimate drivers behind
many scientific questions concerning the Sun --- what is the source of the 11-year sunspot cycle,
what heats the Sun's corona, what accelerates the solar wind, what triggers solar flares,
prominence eruptions, and coronal mass ejections?

Solar granules, bright cells with darker borders, were observed with solar telescopes by the 1860s.
These cellular flows have diameters of about 1Mm, flow speeds of about 3000 m s$^{-1}$, and
lifetimes of about 10 minutes \citep{Bray_etal84}.

The much larger supergranules were first noticed by \cite{Hart54} but named and better characterized
by \cite{Leighton_etal62}.
Supergranules have diameters of about 35 Mm, flow speeds of about 400 m s$^{-1}$, and
lifetimes of about a day \citep{RieutordRincon10}.

Even larger cells, giant cells, were proposed to exist based on theoretical arguments \citep{SimonWeiss68}.
Hints of the possible existence of these cells have been uncovered over the ensuing decades but
detailed observations and characterizations have only come recently \citep{Hathaway_etal13}.
These cells have widths of 100 Mm or more, flow speeds of about 10  m s$^{-1}$, and
lifetimes of several months.

Yet another set of cells, mesogranules, were found by \cite{November_etal81} to be intermediate
in size, flow speeds, and lifetimes between granules and supergranules.

The advent of space based observations, with both continuous coverage and
high spatial and temporal resolution, has enabled many new studies of solar convection.
Data from the Michelson Doppler Investigation (MDI) on the ESA/NASA Solar and
Heliospheric Observatory (SOHO) mission were analyzed by several groups.
\cite{Lawrence_etal99} used a wavelet analysis on Doppler, intensity, and magnetic images.
They noted that granules dominated the radial flows while supergranules dominated the
horizontal flows, but mesogranules were notably weak.
(They also concluded that their wavelet analysis only captured about 30\% of the signal.)
The MDI Doppler data were analyzed and simulated by \cite{Hathaway_etal00}
to obtain photospheric kinetic energy spectra covering spherical harmonic wavenumbers, $\ell$,
from $\ell \sim 1$ to $\ell \sim 3000$.
They concluded that only two spectral features were evident, a peak representing supergranules
at $\ell \sim 120$ and a broad peak representing granules at $\ell > 1000$.
Others \citep[e.g.][]{Rieutord_etal00} have also concluded that mesogranulation is not a
true or discrete scale of solar convection.

Power spectra of the flows at very high wavenumbers were obtained by \cite{Rieutord_etal10} from
high resolution images obtained with the Solar Optical Telescope (SOT) on the {\em Hinode}
spacecraft.
Their horizontal flow spectra show a prominent peak in power at wavenumbers typical for supergranules,
along with a continuum of power out to higher wavenumbers.
(Their correlation tracking method for measuring the horizontal flows limited their measurements to
cells larger than $\sim 2.5$ Mm or $\ell \sim 1800$.)
Their vertical flow spectra (obtained from Doppler velocity measurements) show a prominent
peak in power at wavenumbers typical of granules ($\ell \sim 2500$) with a rapid fall off at
higher wavenumbers.

Here we investigate the Sun's photospheric convection spectrum by analyzing Doppler data from the
HMI instrument.
These data have much higher spatial resolution, and better instrumental characterization, than the MDI data.
The higher resolution data gives better confidence in characterizations of the high-wavenumber
flow structures.
The better characterization of the HMI instrument, its imaging artifacts, and its point-spread-function,
gives better confidence in the characterizations at both ends of the spectrum.

We employ a new Doppler data analysis tool to determine the toroidal component of the
horizontal flows (vortical flows with curl but no divergence).
While the short (compared to the 27-day solar rotation period) lifetimes of supergranules suggest that
solar rotation should not be a big influence on the flows, numerical simulations \citep{Hathaway82}
and observations \citep{DuvallGizon00} indicate that the flows in supergranules have a preferred
sense of vorticity in each hemisphere.
This tendency should be stronger for the larger, slower cellular flows.

In the following section we describe the data and the data preparation.
These data directly measure only the line-of-sight velocity component and only over the visible hemisphere.
We analyze the data with three different analysis methods in an effort to separate the Doppler signal into three
components associated with the full vector velocities in the convective flows.
In order to fully, and accurately, quantify the vector velocities we construct simulated data in which
we know the actual vector velocities over the entire solar surface.
The simulated data are then run through the same analysis procedures,
results are compared with those from the HMI data,
and the simulated data are then revised until a match is obtained.

This data simulation is described in Section 3.
The Doppler velocity spherical harmonic spectrum analysis is described in Section 4.
An analysis that separates the radial flow component from the horizontal flow component
is described in Section 5.
An analysis that separates poloidal flows (diverging horizontal flows) from toroidal flows (horizontal flow vortices)
is described in Section 6.
The full spectrum of motions for all flow components is presented in the concluding section along
with comparisons to the spectrum from a numerical simulation of the smallest length scales, and
to the spectrum of giant cell flow velocities from the correlation tracking method described in \cite{Hathaway_etal13}. 

\section{DATA PREPARATION}

\begin{figure}[ht!]  
\centerline{\includegraphics[width=1.0\columnwidth]{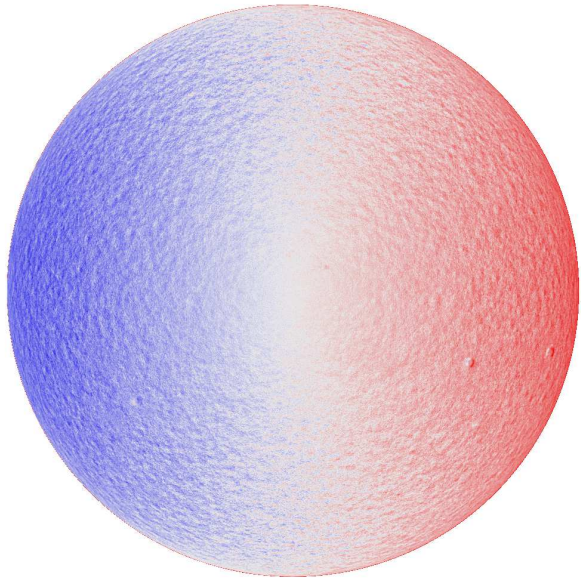}}
\caption[HMI Dopplergram]{Doppler velocity image from the SDO/HMI instrument for November 1, 2012 at 5:00UT.
The line-of-sight velocity signal ranges from -3000 m s$^{-1}$ (dark blue) to +3000 m s$^{-1}$ (dark red).
The signal is dominated by the  $\pm$ 2000 m s$^{-1}$ signal from the Sun's rotation but the mottled pattern
produced by the solar convection is also very apparent (as are ``blemishes'' due to flows around sunspots). 
}
\label{fig:Dopplergram}
\end{figure}

The data used in this study consist of line-of-sight Doppler velocities measured by the HMI
instrument \citep{Scherrer_etal12} on the SDO spacecraft.
These 4096-by-4096 pixel images are obtained every 45 seconds and then averaged over 720 seconds
to remove the 5-minute $p$-mode oscillation signal.
\cite{Schou_etal12} and \cite{Couvidat_etal12} describe the data acquisition and calibration procedures.
The Doppler velocity is calculated from the intensity measured at six different spectral tuning positions
around the wavelength of the selected \ion{Fe}{1} absorption line at 6173 \AA.
An example of a data image is shown in Figure \ref{fig:Dopplergram}.

Our first step is to remove the velocity signal due to the motion of the spacecraft and the velocity artifacts
associated with the instrument itself.
Using the spacecraft velocity relative to the Sun still leaves behind diurnal signals clearly
associated with the spacecraft's geosynchronous orbit.
We found that these variations alter the artifact signal as a function of spacecraft velocity toward or away
from the Sun. The artifacts also change when the instrument focus is changed.
To minimize the impact of these variations on our results, we chose data taken when the spacecraft motion toward
or away from the Sun was less than 300 m s$^{-1}$ (one tenth of the $\pm$ 3000 m s$^{-1}$ daily range).

\begin{figure}[ht!]  
\centerline{\includegraphics[width=1.0\columnwidth]{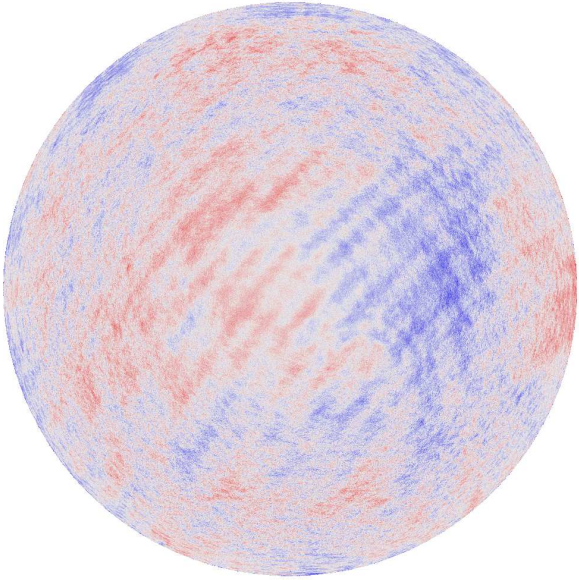}}
\caption[HMI Image Artifacts]{Doppler velocity image showing imaging artifacts from the SDO/HMI instrument for
October 2012 through May 2013.
The line-of-sight velocity signal ranges from -150 m s$^{-1}$ (dark blue) to +150 m s$^{-1}$ (dark red). 
}
\label{fig:Artifacts}
\end{figure}

We extract the imaging artifacts by first removing the large-scale solar velocity signals that are fixed
in the image (e.g. the Doppler signal due to the axisymmetric flows, the gravitational red shift, and the convective blue shift).
We then average over many hundreds of images so that the signal due to the more or less random
convective flows cancel as they evolve and rotate across the visible disk while the artifacts remain fixed.
The artifact image obtained from data acquired
between the dates of October 1, 2012 and May 31, 2013 is shown in Figure \ref{fig:Artifacts}.
These artifacts would introduce spurious features at low spatial wavenumbers in our spectral analyses.

For this study we have analyzed data acquired in November 2012.
After removing the Doppler signals due to the spacecraft motion and the imagining artifacts,
each Doppler image is analyzed using the procedures described in \cite{Hathaway87} and \cite{Hathaway92}
to measure and remove the Doppler signals due to the axisymmetric flows (differential rotation and the
meridional circulation) and the signal due to the convective blue shift.

\begin{figure}[ht!]  
\centerline{\includegraphics[width=1.0\columnwidth]{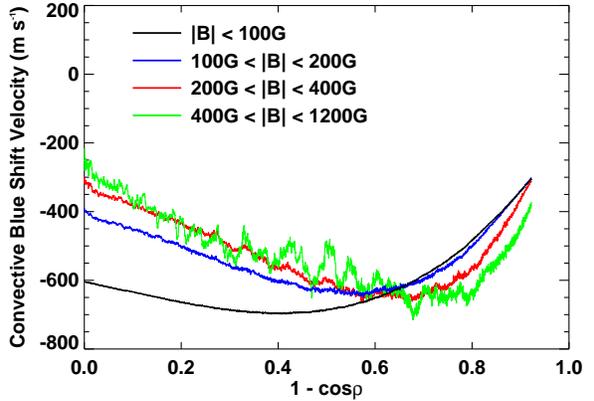}}
\caption[HMI Convective Blue Shift]{The convective blue shift signal from the SDO/HMI instrument for
regions free of magnetic field (black line) and regions with radial magnetic flux density from 100 to 200 G (blue line),
200 to 400 G (red line), and 400 to 1200 G (green line).
}
\label{fig:ConvectiveBlueShift}
\end{figure}

The convective blue shift is an apparent flow toward the observer caused by the correlation between radial
upflows and emergent radiative intensity in the unresolved convective flows (granules).
(The Doppler velocity is an intensity weighted average over the flow structures within a pixel.)
It varies systematically from disk center to limb as the line-of-sight samples different velocity components
and different levels of the solar atmosphere.
The convective blue shift changes substantially in regions where magnetic fields thread through the surface.
Therefore, the convective blue shift signal requires additional corrections.
Figure \ref{fig:ConvectiveBlueShift}  shows the convective blue shift signal for both field free regions
and for regions with varying magnetic flux density.
(Note that while a 636.4 m s$^{-1}$ gravitational red shift has been subtracted from the signal, the original zero
point was arbitrarily set to make the data median match the spacecraft's radial velocity.)
The convective blue shift is substantially altered in magnetic areas.
Uncorrected, this introduces Doppler signals that are not directly related to actual flows on the Sun's surface. 

We fit the convective blue shift signal, $CBS(x)$, to shifted Legendre polynomials (polynomials orthonormal
with uniform weights on the interval $0 \le x \le 1$) with the dependent variable

\begin{equation}
x = 1 - \cos \rho
\end{equation}

\noindent where $\rho$ is the heliocentric angle of a point from disk center.
The first five shifted Legendre polynomials are:

\begin{equation}
P_0^*(x) = 1
\end{equation}

\begin{equation}
P_1^*(x)  = \sqrt{3} (2 x - 1)
\end{equation}

\begin{equation}
P_2^*(x)  = \sqrt{5} (6 x^2 - 6 x + 1)
\end{equation}

\begin{equation}
P_3^*(x)  = \sqrt{7} (20 x^3 - 30 x^2 + 12 x - 1)
\end{equation}

\begin{equation}
P_4^*(x)  = 3 (70 x^4 -140 x^3 + 90 x^2 - 20 x + 1)
\end{equation}

We find that the convective blue shift signal is well fit with these first five polynomials
such that

\begin{equation}
CBS(x) =  \sum_{n=0}^4 C_n P_n^*(x)
\end{equation}

The fit coefficients for the different magnetic flux densities are given in Table \ref{table:CBScoefficients}.

\begin{table}[h!]
\begin{tabular}{| l | rrrrr |}\hline
Flux Density & $C_0$ & $C_1$ & $C_2$ & $C_3$ & $C_4$ \\ \hline
$ < 100$ & -584 & 105 & 92 & 19 & 0 \\
$ [100, 200] $ & -508 & 35 & 113 & 37 & 3 \\
$ [200, 400] $ & -490 & -4 & 123 & 59 & 15 \\
$ [400, 1200] $ & -497 & -35 & 102 & 71 & 33 \\ \hline
\end{tabular}
\caption{Convective blue shift fit coefficients (in units of m s$^{-1}$) for increasing magnetic
flux density (in units of Gauss).}
\label{table:CBScoefficients}
\end{table}

\begin{figure}[ht!]  
\centerline{\includegraphics[width=1.0\columnwidth]{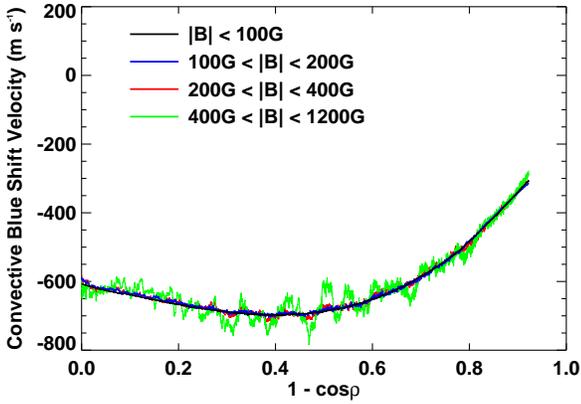}}
\caption[Corrected Convective Blue Shift]{The convective blue shift signal after correcting for magnetic effects.
The four lines represent regions free of magnetic field (black line) and regions with radial magnetic flux density from 100 to 200 G (blue line),
200 to 400 G (red line), and 400 to 1200 G (green line).
}
\label{fig:CBSprofilesCorrected}
\end{figure}

We correct the data for the convective blue shift differences in magnetic regions and find that
a single convective blue shift signal, shown in Figure \ref{fig:CBSprofilesCorrected}, represents
all the data. We then subtract this signal from the data. 

\begin{figure}[ht!]  
\centerline{\includegraphics[width=1.0\columnwidth]{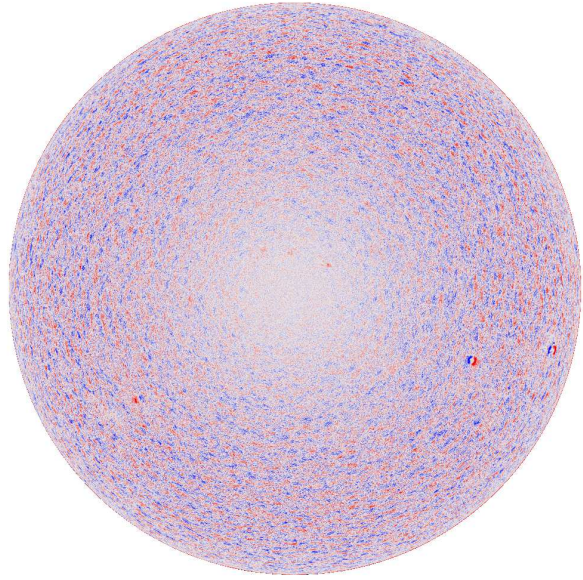}}
\caption[Prepared Dopplergram]{Prepared Doppler velocity image from the SDO/HMI instrument for November 1, 2012 at 5:00UT.
The line-of-sight velocity signal ranges from -1000 m s$^{-1}$ (dark blue) to +1000 m s$^{-1}$ (dark red).
}
\label{fig:PreparedDopplergram}
\end{figure}

After removing these signals (the convective blue shift, the artifacts, the Doppler signal due to differential rotation
and meridional flow)
the Doppler data are dominated by the signal from the line-of-sight velocities in the cellular
flows. A example of these corrected data is shown in Figure \ref{fig:PreparedDopplergram}.

\section{THE DATA SIMULATION}

We produce simulated data (for which we know the full vector velocities over the entire surface of the Sun)
that faithfully reproduce the HMI Doppler data.

We represent the vector velocity field on the surface of a sphere
by a spectrum of poloidal and toroidal modes \citep{Chandrasekhar61} with

\begin{equation}
V_r(\theta,\phi) = \sum_{\ell=0}^{\ell_{max}} \sum_{m=-\ell}^\ell R_\ell^m Y_\ell^m
\end{equation}

\begin{equation}
V_\theta(\theta,\phi) = \sum_{\ell=1}^{\ell_{max}} \sum_{m=-\ell}^\ell \left[
S_\ell^m \frac{\partial Y_\ell^m} {\partial \theta} +
T_\ell^m \frac{1} {\sin\theta} \frac{\partial Y_\ell^m} {\partial \phi} \right]
\end{equation}

\begin{equation}
V_\phi(\theta,\phi) = \sum_{\ell=1}^{\ell_{max}} \sum_{m=-\ell}^\ell \left[ 
S_\ell^m  \frac{1} {\sin\theta} \frac{\partial Y_\ell^m} {\partial \phi} -
T_\ell^m \frac{\partial Y_\ell^m} {\partial \theta} \right]
\end{equation}

\noindent where $Y_\ell^m(\theta,\phi)$ is a spherical harmonic function of degree $\ell$ and azimuthal order $m$,             
$\theta$ is the colatitude measured southward from the north pole, and
$\phi$ is the azimuth measured prograde from the central
meridian. The complex coefficients $R_\ell^m$, $S_\ell^m$, and $T_\ell^m$ are the spectral
coefficients for the radial, poloidal, and toroidal components, respectively.

We simulate the observed line-of-sight velocity by specifying these complex spectral coefficients
for all spectral components up to a maximum spherical harmonic degree,
$\ell_{max} = 4096$.
This is an iterative, trial and error, process in which spectral amplitudes are prescribed,
simulated data are constructed and analyzed, the results are compared to those from the HMI data,
and new spectral coefficients are constructed based on those comparisons.
We start with an estimate of the spectral amplitudes as functions of $\ell$.
The complex coefficients for each azimuthal order, $m$, use these spectral amplitudes but with random
phases for the real and imaginary parts.
We use the same phase for the radial and poloidal spectral coefficients since these two components
are linked through the mass continuity equation ($\nabla \cdot v = 0$) with

\begin{equation}
\ell (\ell + 1) S_\ell^m = \frac{1}{r} \frac{\partial}{\partial r} [r^2 R_\ell^m(r)]
\end{equation}

\noindent We do not link the phase of the toroidal component to the other two.

We calculate the three vector velocity components from this spectrum at an array of points with 4096 points
equi-spaced in $\theta$ and 8192 points equi-spaced in $\phi$.
We then project these vector velocities onto the line-of-sight
at each pixel in a 4096-by-4096 image of the solar disk as seen from 1 AU.
The average Doppler velocity in each pixel is calculated by finding the line-of-sight
projection at a 7-by-7 array of sub-pixels using bi-cubic interpolation from points on the
vector velocity grid and then averaging over those sub-pixels that are on the visible disk.

The line-of-sight velocity at a point $(\theta,\phi)$ is given by

\begin{equation}
\begin{split}
V_{los}(\theta,\phi) =  &V_r(\theta,\phi) \left[ \sin B_0 \cos \theta
+ \cos B_0 \sin \theta \cos \phi \right] +\cr
  &V_\theta(\theta,\phi) \left[ \sin B_0 \sin \theta
- \cos B_0 \cos \theta \cos \phi \right] +\cr
  &V_\phi(\theta,\phi) \left[ \cos B_0 \sin \phi \right]
\end{split}
\label{eqn:LOSprojection}
\end{equation}  
  
\noindent where $B_0$ is the latitude at disk center (or equivalently the tilt
of the Sun's north pole toward the observer) and velocities away from
the observer are taken to be positive.

We also add, at each pixel, a convective blue shift signal that is a function of the heliocentric
angle from disk center.
This signal complicates the extraction of the Doppler signal due to the meridional flow.
If these flows are not properly characterized and removed, they leave behind large-scale
Doppler signals that impact the spectrum at low wavenumbers.
By also including this ``fictious''  signal we faithfully reproduce the full data and analysis
associated with the HMI data.

\begin{figure}[ht!]  
\centerline{\includegraphics[width=1.0\columnwidth]{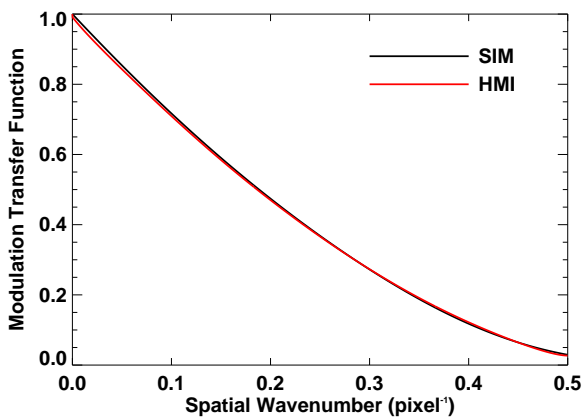}}
\caption[HMI Modulation Transfer Function]{The modulation transfer function (MTF) for the HMI instrument
(red line) and the functional fit used for the simulation (black line). 
}
\label{fig:ModulationTransferFunction}
\end{figure}

A point-spread-function (PSF) was developed to remove stray light from HMI data and is used
here in the data simulation. First, a modulation
transfer function (MTF) was obtained from ground-based calibration data using field stops \citep{Wachter_etal12}
in order to describe the pre-launch instrumental optics. We fit the MTF with an exponential convolved with the
ideal optical transfer function (OTF), often known as a `chat' function. The exponential and `chat' function in
the frequency domain are equivalent to a Lorentzian and Airy function in the spatial domain.
Post-launch data including solar aureole, lunar eclipse and Venus transit events were used to
evaluate how well the PSF was able to reproduce the observed scattering.

The final value of gamma (the wavenumber exponent) in the exponential function was determined through least squares fitting
of the transit of Venus data on 2012.06.05 from the side camera data.
The final, non-ideal form of the MTF and subsequent PSF was further modified for the following reasons.
Lunar eclipse images, where a large portion of the solar image is obscured by the lunar disk, enabled
a measure of the large-scale scattering. One such eclipse occurred on 2010.10.07 and the HMI continuum intensity
filtergram shows a light level of 0.34\% of the disk-center continuum intensity for a position 200 pixels onto the
lunar disk. We found that the light level tended towards a constant far away instead of continually decreasing
with increasing distance from the solar limb. This motivated an additional term to be added to the PSF
to fit the tail of the distribution. If we only considered light scattered from 10$^{\prime\prime}$ away
then the additional term would not be necessary, meaning that using only the Venus transit data was not
sufficient in order to emulate the large-scale scattering/dust on the optics.
(Note that the PSF we developed differs from previous stray light removal efforts since we do not use
a single Gaussian or sum of Gaussians as the central mathematical component.)

Deconvolved images were compared to the originals
and it was found that the minimum intensity of a sunspot umbrae decreased from being 5.5\% of the nearby quiet-Sun
continuum intensity in the original image to being 3.3\% in the deconvolved image while the granulation contrast doubled,
with a standard deviation of the intensity in the quiet-Sun being 3.7\% of the average in the original image and
7.2\% of the average in the deconvolved image.

In our data simulation we average the line-of-sight velocity over all sub-pixels within a pixel that fall on the visible disk of the Sun
and then convolve that image with the PSF found for the HMI instrument.
This convolution is implemented by multiplying the FFT of the image with the FFT of the PSF
(the MTF) and then performing an inverse FFT on the result.
The MTF found for HMI is shown in Figure \ref{fig:ModulationTransferFunction} along with
the functional fit (which is what we actually use) given by 

\begin{equation}
MTF(k) = \exp(-k^5) (1 - k/2)^3 \sin(k \pi/2)/(k \pi/2)
\end{equation}

\begin{figure}[ht!]  
\centerline{\includegraphics[width=1.0\columnwidth]{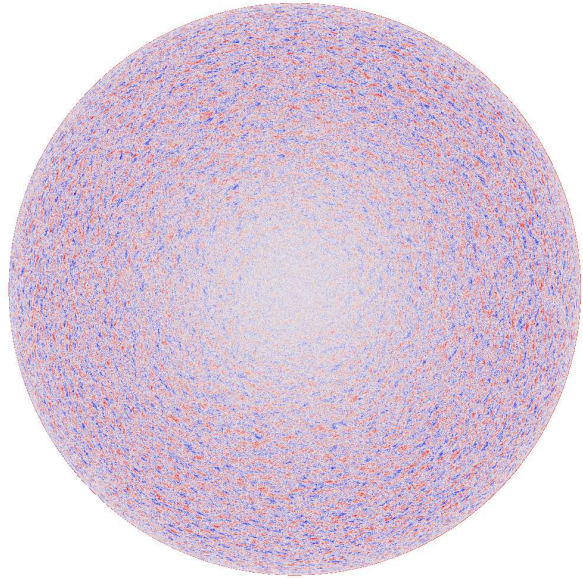}}
\caption[Simulated Data]{Simulated Doppler velocity image.
The line-of-sight velocity signal ranges from -1000 m s$^{-1}$ (dark blue) to +1000 m s$^{-1}$ (dark red).
}
\label{fig:SimulatedData}
\end{figure}

An example of a Doppler image from the data simulation after the signals due to the axisymmetric flows and the
convective blue shift are removed is shown in Figure \ref{fig:SimulatedData}.

\section{DOPPLER VELOCITY SPECTRUM}

We calculate Doppler velocity spectra following \cite{Hathaway_etal00}.
We remove active regions in the HMI data by locating pixels where the magnetic flux density exceeds 1200 G
and then masking out those pixels along with all adjacent pixels within a 10 pixel radius.
The full disk HMI data, its mask, and the simulated data are projected onto maps in heliographic longitude and latitude.
The mask edges are smoothed by convolving the mask with a cosine bell having a radius of 10 pixels.
The masked data is then projected onto spherical harmonics by taking Fourier transforms in longitude and
Legendre transforms in latitude to obtain complex spectral coefficients, $A_\ell^m$, that reproduce the
masked data when multiplied by the spherical harmonics and summed over $\ell$ and $m$.

\begin{figure}[ht!]  
\centerline{\includegraphics[width=1.0\columnwidth]{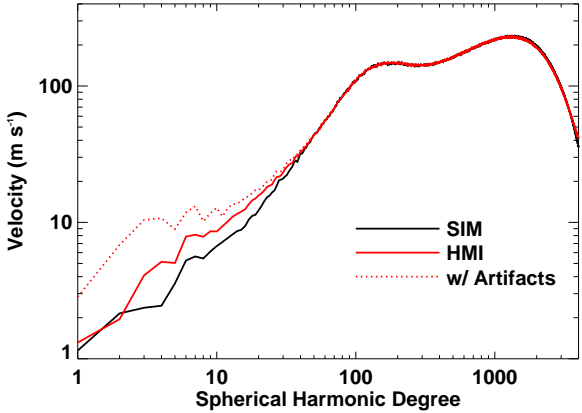}}
\caption[Doppler Velocity Spectra]{The Doppler velocity spectra for the simulated data
(black line) and the HMI data (red line). The dotted red line shows the spectrum obtained
from the HMI data when the artifacts are not removed. 
}
\label{fig:DopplerVelocitySpectra}
\end{figure}

The HMI data are comprised of 59 Doppler images acquired during November 2012 at the hourly intervals
when the spacecraft motion toward or away from the Sun was less than 300 m s$^{-1}$.
The simulated data are comprised of 18 Doppler images constructed with different sets of random numbers
for the spectral phases.
The average velocity spectra are shown in Figure \ref{fig:DopplerVelocitySpectra}.

These spectral values are calculated by taking the square root of the wavenumber times the
velocity power per wavenumber:

\begin{equation}
V(\ell) = \sqrt{\ell \sum_{m=-\ell}^\ell |A_\ell^m|^2}
\end{equation}

The Doppler velocity spectrum rises to a peak at $\ell \sim 120$ (consistent with typical
supergranules with diameters of $\sim35$ Mm), drops slightly and then rises again to a
second peak at  $\ell \sim 3500$ (consistent with granules with diameters of $\sim1.2$ Mm).

Note that this spectrum is for the Doppler velocities.
It includes foreshortening at the limb, which limits the signal at high wavenumbers,
and it includes line-of-sight effects, which exclude horizontal velocities at disk center
and only include one horizontal velocity component across the rest of the disk.
In spite of these caveats, it is nonetheless clear that the spectrum is continuous with
only two primary features --- peaks representative of granules and supergranules.

While cells of all sizes are included, there isn't a spectral feature that indicates
a preference for mesogranules of any particular size.
The bottom of the spectral dip at $\ell \sim 400$ represents cells with diameters
of $\sim 10$ Mm --- a quoted size for mesogranules.

We have forced the simulation spectrum to drop smoothly at low wavenumbers.
While this leaves an apparent excess of power in the HMI data at wavenumbers
below 30, we attribute this signal excess to a less than perfect removal of the
image artifacts. This is supported by the low wavenumber feature seen in the data
when the artifacts are not removed (dotted red line in Figure \ref{fig:DopplerVelocitySpectra}).

\section{RADIAL FLOW ANALYSIS}

The radial flow component represented by $R_\ell^m$ can be separated from the horizontal flow
components using the method described by \cite{Hathaway_etal02A}.
The Doppler velocity due to the horizontal flows drops to zero at disk center, where
all horizontal velocities are transverse to the line-of-sight, while the radial velocity
has its strongest contribution to the Doppler signal at disk center.

The line-of-sight velocity is given by

\begin{equation}
V_{los}(\theta,\phi) =  V_r(\theta,\phi)  \cos \rho + V_{h1}(\theta,\phi) \sin \rho
\end{equation}

\noindent where $V_{h1}$ is the component of the horizontal flow velocity aligned with
the radius vector from disk center, and $\rho$ is the heliocentric angle from disk center.
The orthogonal component of the horizontal flow is transverse to the line-of-sight and does
not contribute to the Doppler signal.
Squaring the Doppler velocity and averaging it in annuli about disk center at a given $\rho$ gives

\begin{equation}
\begin{split}
\overline{V_{los}^2(\rho)} = &\overline{V_r^2} \cos^2 \rho +  \overline{V_{h1}^2} \sin^2 \rho +\cr
&2\overline{V_r V_{h1}} \cos \rho \sin \rho
\end{split}
\end{equation}
where the overbars represent the averages.
We expect the last term in this equation to be negligible.
There is no physical or geometric reason that $V_r V_{h1}$ should have a nonzero average on an annulus. 
For example, the up flow at the center of a cell should
produce horizontal flows away from disk center on one side and toward disk center on the other -- giving a
net zero average. Dropping this term, we rewrite the equation as

\begin{equation}
\overline{V_{los}^2(\rho)} = \overline{V_r^2} + 
[\overline{V_{h1}^2} - \overline{V_r^2}] \sin^2 \rho
\end{equation}

This suggests a method for separating the radial flow from the horizontal flow component in the
Doppler velocity signal. If we plot the average Doppler velocity squared as a function of $\sin^2 \rho$,
then the y-axis intercept gives the radial velocity squared and the slope gives the difference between
the square of the horizontal component and the square of the radial component.

We need to know the contributions to the Doppler signal from the radial and
horizontal flows as functions of wavenumber.
To do so we reconstruct the Doppler signal from the spherical harmonic spectral coefficients, $A_\ell^m$,
multiplied by spectral filters to isolate Doppler features of different sizes.
We chose a set of Gaussian filters of width $\Delta \ell = 64$ centered on multiples, n, of $\ell=128$
from $\ell =0$ to $\ell = 4096$ with

\begin{equation}
Filter_n(\ell) = \exp[-(\ell - n 128)^2/(2 \Delta \ell^2)]
\end{equation}

\begin{figure}[ht!]  
\centerline{\includegraphics[width=1.0\columnwidth]{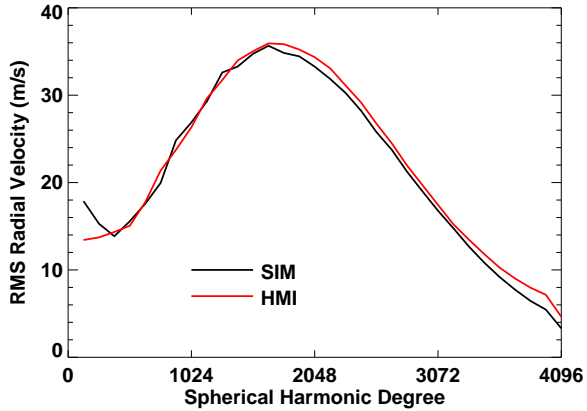}}
\caption[Radial Doppler Velocity]{The radial flow component of the Doppler velocity for the simulated data
(black line) and the HMI data (red line). 
}
\label{fig:RadialDopplerVelocity}
\end{figure}

\begin{figure}[ht!]  
\centerline{\includegraphics[width=1.0\columnwidth]{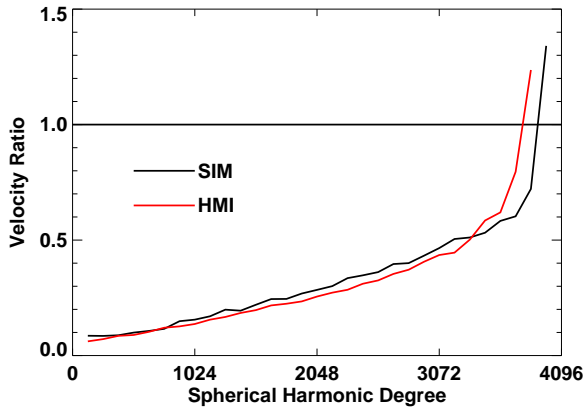}}
\caption[Radial/Horizontal Velocity Ratio]{The ratio of the radial flow component to the horizontal
flow component of the Doppler velocity for the simulated data
(black line) and the HMI data (red line). 
}
\label{fig:RadialHorizontalVelocityRatio}
\end{figure}

The average Doppler velocity squared for each of these 33 filtered Doppler images is fit to a second
order polynomial in $\sin^2 \rho$ (foreshortening away from disk center introduces curvature to these data).
The y-axis intercept yields the radial flow velocity squared while the slope yields the difference
between the squared radial  and squared horizontal
flow velocities. The radial flow components are shown in Figure \ref{fig:RadialDopplerVelocity}
as functions of $\ell$ for both the HMI and the simulated data.
The ratios of the radial flow component to the horizontal flow component are shown in
Figure \ref{fig:RadialHorizontalVelocityRatio} as functions of $\ell$ for both the HMI and the simulated data.

Note that this analysis compares the radial flow velocity to just one component of the horizontal
flow velocity. While it is clear that the radial flow is very weak at low wavenumbers and becomes
comparable to the horizontal flows at large wavenumbers, a quantitative value is best obtained
from the vector velocities in the simulation.

The simulation has radial flow velocities that are just 3\% of the total
flow velocity at the lowest wavenumbers ($\ell \sim 10$). This ratio rises to 6\%
at supergranule wavenumbers ($\ell \sim 120$) and to 45\% at granule wavenumbers ($\ell \sim 4000$).

\section{TOROIDAL/POLOIDAL FLOW ANALYSIS}

We have devised a new analysis technique to separate the toroidal flow component from the poloidal
flow component.
We construct two velocity gradient images from the prepared data by calculating the pixel-wise gradient
of the Doppler velocity (a scalar) at each point in the image and finding the projection of that gradient in the disk-radial direction
(from disk center toward the limb) to produce one image (the Grad$_\rho$ image) and the
projection of the gradient in the tangential direction (clockwise about disk center) to produce the
second image (the Grad$_\Theta$ image).

A poloidal flow away from a convection cell center
will contribute to the Grad$_\rho$ image but not to the Grad$_\Theta$ image
along lines that pass radially and tangentially through the cell center.
Likewise, a toroidal flow around the center of a vortex anywhere on the disk will contribute
to the Grad$_\Theta$ image but not to the Grad$_\rho$ image along these same lines.
Unfortunately, a radial flow will contribute to both images and both poloidal and toroidal flows
will contribute to both images away from the lines that pass through cell centers in the
radial and tangential directions.
While the separation into poloidal and toroidal components is not clean and simple,
the data simulation allows us to nonetheless quantify the contributions of these flows
to the Grad$_\rho$ and Grad$_\Theta$ images.

We produce these Grad$_\rho$ and Grad$_\Theta$ images for both the HMI and the simulated data
and then find their spherical harmonic spectral coefficients using the same procedure used
for the Doppler velocity spectrum in Section 4.

\begin{figure}[ht!]  
\centerline{\includegraphics[width=1.0\columnwidth]{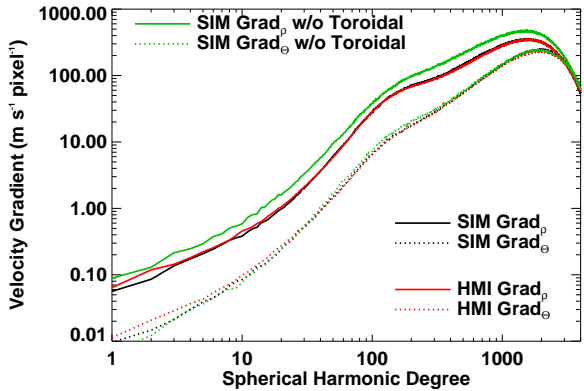}}
\caption[Toroidal Poloidal Velocity Spectra]{The spectra from the Grad$_\rho$ images(solid lines)
and Grad$_\Theta$ images (dotted lines)
for the simulated data with both poloidal and toroidal flows (black lines), the HMI data (red lines), and the
simulated data in which the horizontal flows are purely poloidal without a toroidal component (green lines). 
}
\label{fig:ToroidalPoloidalVelocitySpectra}
\end{figure}

The results are shown in Figure \ref{fig:ToroidalPoloidalVelocitySpectra}.
We illustrate the sensitivity of the method to poloidal and toroidal flows using two simulations.
Both simulations have horizontal flows that produce the spectral match shown in
Figure \ref{fig:DopplerVelocitySpectra} and radial flows that produce the spectral matches shown in
Figures \ref{fig:RadialDopplerVelocity} and \ref{fig:RadialHorizontalVelocityRatio}.
One simulation, shown in black in Figure \ref{fig:ToroidalPoloidalVelocitySpectra}, has both poloidal
and toroidal flows adjusted to match the HMI Grad$_\rho$and Grad$_\Theta$ spectra (shown in red).
The second simulation (shown in green), has only poloidal flows without a toroidal flow component.
The excess velocity at all wavenumbers in the Grad$_\rho$ image spectrum from this simulation indicates that a substantial
fraction of the observed horizontal flows must be toroidal.
The deficient velocity at wavenumbers below $\ell \sim 30$ in the Grad$_\Theta$ image spectrum from this
simulation suggests that the toroidal flows dominate at these lower wavenumbers.

\section{CONCLUSIONS}

\begin{figure}[ht!]  
\centerline{\includegraphics[width=1.0\columnwidth]{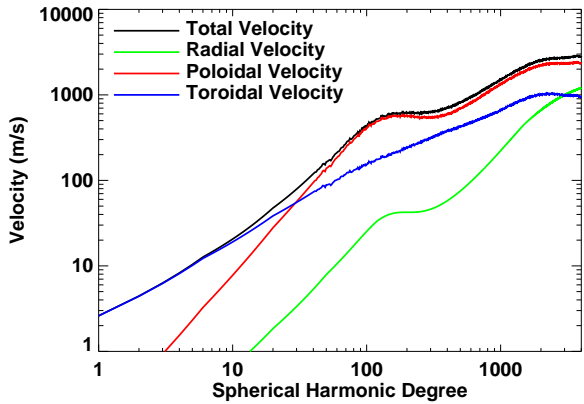}}
\caption[Simulation Velocity Spectrum]{The total velocity spectrum (black line) from
the data simulation. The contribution from the poloidal flow is shown in red.
The contribution from the toroidal flow is shown in blue. 
The contribution from the radial flow is shown in green.  
}
\label{fig:SimulationVelocitySpectrum}
\end{figure}

As a result of these analyses, we have simulated flows that closely reproduce the analysis
results from the HMI data seen in the spectral matches in the previous sections.
The simulations require prescribing all three spectral coefficients (radial, poloidal, and toroidal)
for all wavenumbers from 1 to 4096.
The total velocity spectrum and its three components are plotted in Figure \ref{fig:SimulationVelocitySpectrum}
as functions of spherical harmonic degree $\ell$.
Note that these spectra are from the vector velocities over the entire solar surface and that
they are not impacted by the foreshortening or line-of-sight projection effects that impact the spectra from the Doppler images.

As with the Doppler Velocity spectrum, we see that the total velocity spectrum rises to the peak at $\ell \sim 120$,
consistent with supergranules, drops slightly at higher wavenumbers and then rises to the
second peak at $\ell \sim 3500$, consistent with granules.
The two velocity components coupled by the mass continuity equation, the radial and poloidal
components, have pronounced peaks consistent with supergranules while the toroidal component
does not.

None of the components show any spectral features indicating a preferred cell size at the
lowest wavenumbers representing giant cells or at the intermediate wavenumbers representing
mesogranules.

We conclude that, in general, the convection spectrum is a rising spectrum with velocities increasing
linearly (slope of $\sim 1$ on this log-log plot) with wavenumber $\ell$ to an ultimate peak at wavenumbers of $\ell \sim 3500$
consistent with the photospheric granulation.
However, this linear rise is punctuated by a peak at $\ell \sim 120$ consistent with supergranules.

The origin of this supergranule peak is still uncertain.
\cite{SimonLeighton64} suggested that cells the size of supergranules are driven by the
changes in mean molecular weight at depths where helium becomes ionized
(depths of $\sim 7$ Mm for the first ionization of He and  $\sim 30$ Mm for the second
ionization of He).
\cite{November_etal81} isolated mesogranule-sized cells and suggested that mesogranules
were driven by the first ionization of He and supergranules by the second ionization of He.
\cite{Rast03} suggested that cells the size of mesogranules and supergranules would be
driven by the collective interactions between downdrafts at cell boundaries.
However, radiative-MHD simulations of convection in the outer 20 Mm of the Sun by
\cite{Stein_etal11} include all of these processes, but do not yield any dominant cells sizes
in the photosphere other than granules.
(These simulation do, however, indicate that the smaller structures disappear in deeper layers,
leaving velocity structures with sizes proportional to the depth.)

The relative strength of the radial flow velocity to the total flow velocity increases from 3\%
at the lowest wavenumbers to nearly 50\% at the highest wavenumbers.
This is in rough agreement with the results found with the lower resolution MDI data by \cite{Hathaway_etal02A},
but extends the results to much higher wavenumbers, and slightly lowers the radial flow velocities
at the lowest wavenumbers (where imaging artifacts were removed from the HMI data).

The toroidal flow component dominates at low wavenumbers and becomes comparable
to the poloidal component at $\ell \sim 30$, where velocities of $\sim 70$ m s$^{-1}$ and
cell sizes of $\sim 150$ Mm give a characteristic time scale of $\sim 24$ days --- comparable to
the Sun's rotation period.

The spectrum shown in Figure \ref{fig:SimulationVelocitySpectrum} is derived from
Doppler velocity data obtained with HMI.
This spectrum can be compared to spectra obtained by other means.

\begin{figure}[ht!]  
\centerline{\includegraphics[width=1.0\columnwidth]{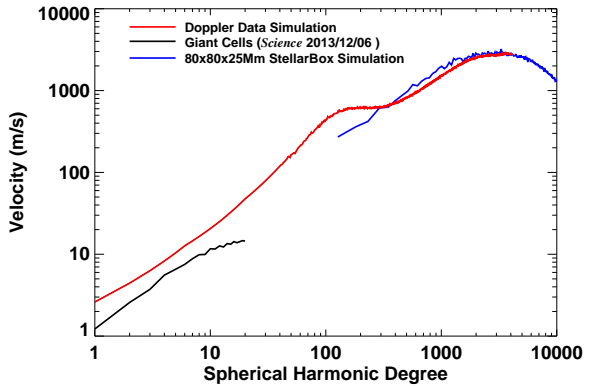}}
\caption[Composite Velocity Spectrum]{The total velocity spectrum (red line) from
the data simulation is plotted for comparison with the spectrum from the vector
velocities found for giant cells \citep{Hathaway_etal13} and the spectrum from
the vector velocities in the  StellarBox simulation of solar granulation.
}
\label{fig:CompositeVelocitySpectrum}
\end{figure}

\cite{Hathaway_etal13} measured the giant cell sized flows by a local correlation tracking
method on the Doppler velocity features (largely supergranules).
This analysis produced maps of the horizontal vector velocities for all solar
longitudes and most solar latitudes (the extreme polar regions were inaccessible).
The velocity spectrum averaged over 34 solar rotations from May 2010 through
November 2012 is shown in Figure \ref{fig:CompositeVelocitySpectrum} by the
black line at low wavenumbers.
The giant cell flow velocities are somewhat smaller that those in our simulation,
but the slope agrees very well.
Our data simulation may overestimate the velocity at these low wavenumbers
by having to fit to HMI data that is still impacted by the presence of imaging artifacts.
On the other hand, the giant cell flow velocities found by \cite{Hathaway_etal13} 
may be underestimated due to averaging the moving and evolving flows at each point over the
10-13 days that point is visible in HMI data.

Numerical simulations for the three-dimensional radiative magnetohydrodynamics
have progressed to the point that they now reproduce the appearance, the time evolution,
and the emergent radiation associated with granules \citep{Kitiashvili_etal12}.
We have compared our Doppler velocity spectrum with the spectrum calculated from the
three-dimensional radiative hydrodynamics simulations obtained using the StellarBox code
\citep{Wray_etal15}, which takes into account the effects of ionization, sub-grid-scale turbulence,
internal structure and chemical composition.
For analysis we used a fully developed hydrodynamic run in a slab 80 by 80 Mm in the horizontal,
25 Mm in depth below the photosphere, and 1 Mm of the atmospheric layer with the spatial grid resolution of 100 km.
The photospheric velocity spectrum from this simulation (shown in Figure \ref{fig:CompositeVelocitySpectrum}
 by the blue line at high wavenumbers) is in very good agreement with the HMI Doppler velocity spectrum
 in the range of the spherical harmonic degree from 200 to 4000.
 However, the supergranulation bump is missing in these numerical simulations (as it was in
 the simulations of \cite{Stein_etal11}).
 
 Understanding of the origin of supergranulation is still a significant challenge in solar physics.
 Origins based on the ionization of helium \citep{SimonLeighton64} or the collective interactions between downdrafts
 \citep{Rast03} now seem unlikely based on numerical simulations.
 The supergranules are imbedded in the Sun's surface shear layer (which extends to a depth of $\sim 50$ Mm)
 and are characterized by the magnetic network that forms around the periphery of each cell.
 Perhaps their origin lies with the influence of the Sun's rotation or magnetic field on these convective flows.

\acknowledgements
The authors were supported by a grant from the NASA Living with a Star Program to Ames Research Center.
The HMI data used are courtesy of the NASA/SDO and the HMI science teams.
We gratefully acknowledge many useful discussions with Nagi Mansour, Alan Wray, Alexander Kosovichev, and Philip Scherrer.



\end{document}